\documentclass[aps,prl,showpacs,superscriptaddress,twocolumn]{revtex4}

\usepackage{graphicx}
\usepackage{times}
\usepackage{color}
\newcommand{\ket}[1]{{\left|{#1}\right\rangle}}
\newcommand{\bra}[1]{{\left\langle{#1}\right|}}
\newcommand{\tr}[1]{\textnormal{tr}{\left\{#1\right\}}}
\newcommand{\heading}[1]{\par\textbf{#1}\quad\ignorespaces}
\newcommand{\repr}{\mathrel{\widehat{=}}}
\newcommand{\I}{\mathrm{i}}
\newcommand{\Exp}[1]{\mathrm{e}^{\mbox{\footnotesize$#1$}}}
\newcommand{\expect}[1]{\langle{#1}\rangle}
\newcommand{\bexpect}[1]{\bigl\langle{#1}\bigr\rangle}

\begin{document}
\title{Experimental detection of entanglement with optimal-witness families}
\author{Jibo Dai}
\affiliation{Data Storage Institute, Agency for Science, Technology and
  Research, 5 Engineering Drive 1, Singapore 117608, Singapore} 
\affiliation{Centre for Quantum Technologies, National University of
  Singapore, 3 Science Drive 2, Singapore 117543, Singapore} 
\author{Yink Loong Len}
\affiliation{Data Storage Institute, Agency for Science, Technology and
  Research, 5 Engineering Drive 1, Singapore 117608, Singapore} 
\affiliation{Centre for Quantum Technologies, National University of
  Singapore, 3 Science Drive 2, Singapore 117543, Singapore} 
\affiliation{Department of Physics, National University of Singapore, 2
  Science Drive 3, Singapore 117542, Singapore} 
\author{Yong Siah Teo} 
\affiliation{Centre for Quantum Technologies, National University of
  Singapore, 3 Science Drive 2, Singapore 117543, Singapore} 
\affiliation{Department of Optics, Palack{\'y} University, 17.~listopadu 12,
  77146 Olomouc, Czech Republic} 
\author{Berthold-Georg Englert}
\affiliation{Centre for Quantum Technologies, National University of
  Singapore, 3 Science Drive 2, Singapore 117543, Singapore} 
\affiliation{Department of Physics, National University of Singapore, 2
  Science Drive 3, Singapore 117542, Singapore} 
\author{Leonid A. Krivitsky}
\affiliation{Data Storage Institute, Agency for Science, Technology and 
  Research, 5 Engineering Drive 1, Singapore 117608, Singapore}

\date[]{Posted on the arXiv on 24 February 2014}

\begin{abstract}
We report an experiment in which one determines, with least tomographic effort, 
whether an unknown two-photon polarization state is entangled or separable. 
The method measures whole families of optimal entanglement witnesses.
We introduce adaptive measurement schemes that greatly speed up the
entanglement detection.  
The experiments are performed on states of different ranks, 
and we find good agreement with results from computer simulations. 
\end{abstract}


\maketitle

\heading{Introduction}
Entangled states play an important role in the manipulation of quantum
information, be it for present-day quantum key distribution or future quantum
computation.
One may need to verify if a certain quantum state---perhaps emitted by a
source of quantum-information carriers or obtained as the output of a quantum
computation---is entangled or not. 
For this purpose, the expectation value of an \emph{entanglement witness}
is telling: 
The state is surely entangled if a negative value is obtained.
A positive value, however, is inconclusive---the unknown state could be
entangled or separable, the witness cannot tell, but other witnesses might be
able to. 
The concept of a witness was first used by M., P., and R.~Horodecki
\cite{Horodeccy:1996}, and the term ``witness'' was introduced by Terhal
\cite{Terhal:2000};
for reviews that cover all important aspects of entanglement witnesses, see
Refs.~\cite{Guehne+Toth:2009,Horodeccy:2009}. 

How many witnesses, then, does one need to measure until a conclusion is
reached?
The answer to this question is given in Ref.~\cite{Zhu+2:2010}:
If one solely relies on the expectation values of the witnesses one by one,
one may never get a conclusive answer; if, however, the expectation values of
suitably chosen witnesses are jointly used for an estimation of the quantum
state, ${D^2-1}$ witnesses suffice for a $D$-dimensional quantum system.

The number can be further reduced by exploiting all the information 
gathered when determining the expectation values of the witnesses measured
in succession. 
In the case of a two-qubit state (${D=4}$), for instance, one never has to
measure more than six witnesses, rather than ${15=4^2-1}$.
This can be demonstrated by an experiment such as the one proposed in
Ref.~\cite{Zhu+2:2010}; we are here reporting its laboratory realization.

\heading{Witnesses and witness families}
A hermitian observable $W$ is an entanglement witness if
${\tr{\rho_{\mathrm{sep}}W}\geq0}$ for \emph{all} separable states
$\rho_{\mathrm{sep}}$
and ${\tr{\rho_{\mathrm{ent}}W}<0}$ for at least one entangled state
$\rho_{\mathrm{ent}}$. 
For each entangled state, there are some witnesses that detect it (``$<0$''), 
but many other witnesses will give an inconclusive result (``$\geq0$'').

We shall concern ourselves with two-qubit systems---in the experiment, they
are polarization qubits of a down-converted photon pair---and focus on optimal
  decomposable witnesses \cite{Lewenstein+3:2000} of the form 
${W=\bigl(\ket{\mathrm{w}}\bra{\mathrm{w}}\bigl)^{\mathrm{T}_2}}$,
where $\ket{\mathrm{w}}$ is the ket of an entangled pure two-qubit state, and
$^{\mathrm{T}_2}$ denotes the partial transposition on the second qubit. 
This witness is optimal in the sense that no other witness can detect
some entangled states in addition to the states already detected by $W$.
The generic example is
${\ket{\mathrm{w}}=\ket{00}\cos(\frac{1}{2}\alpha)
+\ket{11}\sin(\frac{1}{2}\alpha)}$ with $\sin\alpha\neq0$, on which all other
$\ket{\mathrm{w}}$s can be mapped by local unitary transformations.   

For any $\alpha$, the eigenkets of the resulting witness,
\begin{eqnarray}\label{eq:Walfa}
  W^{(\alpha)}&=& \ket{00}\frac{1+\cos\alpha}{2}\bra{00}
                  +\ket{11}\frac{1-\cos\alpha}{2}\bra{11}\nonumber\\
               &&\mbox{}  
                  +\ket{\Psi_+}\frac{\sin\alpha}{2}\bra{\Psi_+}
                  -\ket{\Psi_-}\frac{\sin\alpha}{2}\bra{\Psi_-}\,,
\end{eqnarray}
are the same: the two
product kets $\ket{00}$ and $\ket{11}$ as well as the two Bell kets
${\ket{\Psi_\pm}=\left(\ket{01}\pm\ket{10}\right)/\sqrt{2}}$. 
This whole family of entanglement witnesses can, therefore, be
measured by the projective measurement of their common eigenstates---the
\emph{witness basis} of the family.
For a separable state, ${\tr{\rho_{\mathrm{sep}}W^{(\alpha)}}\geq0}$ 
for all $\alpha$, and this requirement implies the witness-family criterion
\cite{Zhu+2:2010}
\begin{equation}
\mathcal{S} \equiv 4f_1f_2-\left(f_3-f_4\right)^2\geq0\,.
\label{eq:witcriterion}
\end{equation}
Here, $f_1$ and $f_2$ are the probabilities for the two product
states, and $f_3$ and $f_4$ are those for the Bell states; we estimate these
probabilities from the observed frequencies. 
Consequently, once the frequency data are obtained from the measurement, 
a negative value of $\mathcal{S}$ reveals that the unknown state
$\rho_{\mathrm{true}}$ is entangled. 

Three remarks are in order.
(i) The expectation value of one witness $W^{(\alpha)}$ is a linear
function of the $f_j$s, whereas $\mathcal{S}$ is a quadratic function.
This is reminiscent of, yet different from, the ``nonlinear entanglement
witness'' of Ref.~\cite{Kotowski+2:2010}, which requires a joint measurement
on two copies of the unknown state. 
Our \emph{witness-family measurement} uses only one copy at a time.
(ii)  Also, the nonlinear witnesses of
Ref.~\cite{Guhne+1:2006} are quite different; 
their evaluation requires complete or almost complete knowledge of the state.
(iii) The witnesses $W^{(\alpha)}$ can also be measured by other
schemes, such as that of Barbieri \textit{et al.} \cite{Barbieri+5:2003} who
extracted the expectation value of $W^{(\pi/2)}$ from local measurements that
examine the two qubits individually.  
By contrast, we perform a joint measurement on both qubits, 
which implements the projective measurement in the eigenbasis of
$W^{(\alpha)}$ and so realizes
the most direct measurement of the witness. See the Appendix
for more details about (ii) and (iii).

Our witness-family measurement provides estimates for
\emph{three}  two-qubit observables (four $f_j$s with unit sum), 
whereas the expectation value of a single witness is only one number. 
As discussed in Ref.~\cite{Zhu+2:2010}, this can be exploited for
quantum-state reconstruction after measuring six witness families, related to
each other by the six local unitary transformations of Table~\ref{tbl:six}: 
Each witness family provides the expectation values of one (of six)
single-qubit observables and of two (of nine) two-qubit observables.
Therefore, a measurement of all six witness families constitutes an informationally
complete (IC) measurement for full tomography of an unknown two-qubit state; 
thereby, all six single-qubit parameters and six two-qubit parameters are
obtained once, while three two-qubit parameters are determined twice.    
This offers the possibility of measuring an IC set of witness families such
that, if all families give an inconclusive result ($\mathcal{S}\geq0$), a full
state estimation can be performed for identifying $\rho_{\mathrm{true}}$. 
With $\rho_{\mathrm{true}}$ then at hand, its separability can be determined
straightforwardly by, for example, checking the Peres-Horodecki criterion
\cite{Peres:1996,Horodeccy:1996}.

\begin{table}
\caption{\label{tbl:six}%
The six witness families that enable full tomography of the two-qubit state. 
The single-qubit unitary operators $U_1$ and $U_2$ transform the first family
into the other five families.
The Pauli operator $X$ permutes $\ket{0}$ and $\ket{1}$; the
Clifford operator $C$ permutes the three Pauli operators cyclically.} 
\centering
    \begin{tabular}{c@{\qquad}c@{\quad}c}
    \hline \hline
Family & $U_1$ & $U_2$ \\  \hline
     1 & $\mathbf{1}$ & $\mathbf{1}$\\ 
     2 & $\mathbf{1}$ & $X$\\
     3 & $C^\dagger$ & $C$\\  
    \hline \hline
    \end{tabular}
\qquad
    \begin{tabular}{c@{\qquad}c@{\quad}c}
    \hline \hline
Family & $U_1$ & $U_2$ \\  \hline
     4 & $C^\dagger$ & $XC$\\ 
     5 & $C$ & $C^\dagger$\\
     6 & $C$ & $XC^\dagger$\\  
    \hline \hline
    \end{tabular}
\end{table}

\heading{Scheme~A: Random sequence}
Since $\rho_{\mathrm{true}}$ is unknown, there is no preference for a
particular one of the six families to start with. 
Hence, one starts with a randomly chosen family 
and checks the inequality (\ref{eq:witcriterion}). 
If the result is inconclusive, one then chooses the next family at random from
the remaining five families, and so forth until a conclusive result is
obtained.  
If all six families give inconclusive results, $\rho_{\mathrm{true}}$ is estimated
from the data to establish if it is entangled or separable.

\heading{Scheme B: Adaptive measurements}
Alternatively, we can perform the witness-family measurements in an adaptive
manner: 
We choose the next family to be measured in accordance with the data
obtained from previous measurements.  
Each time a witness family is measured, a set of four frequencies is obtained
and these \emph{informationally incomplete} data are used to partially
estimate $\rho_{\mathrm{true}}$ by, for example, jointly maximizing the 
likelihood and the entropy---the MLME strategy of
Ref.~\cite{Teo+4:2011,Teo+4:2012}.  
The MLME estimators $\rho_{\textsc{mlme}}$ tend to be highly-mixed states and are
thus hard to detect by entanglement witnesses; therefore, if the measurement
of a witness family detects the entanglement of the MLME estimator by the
criterion (\ref{eq:witcriterion}), measuring that family has a good chance of
detecting the entanglement of $\rho_{\mathrm{true}}$. 
The value of $\mathcal{S}$ is used for comparing the
unmeasured witness families with the MLME estimator, with $f_j$ replaced by the
$j$th MLME probability $\tr{\rho_{\textsc{mlme}}\,U^\dagger\Pi_jU}$ where
${U=U_1\otimes U_2}$ is one of the six unitary operators of Table~\ref{tbl:six},
and $\Pi_j$ projects to the $j$th ket in the witness bases.
The family that gives the smallest value of $\mathcal{S}$ is measured next;
this judicious choice of family reduces the average number of witness families
that need to be measured before the entanglement is detected.
Instead of fixing the above six families, based on the MLME
estimator, one can also choose from all thinkable families in each step. 
However, it is not worth the trouble as such optimization hardly
improves the entanglement detection; see the Appendix
for details.

\heading{Scheme C: Maximum-likelihood set} 
All state estimators, including $\rho_{\textsc{mlme}}$, that maximize the
likelihood compose a convex set, the \emph{maximum-likelihood (ML) set}. 
They all give the same estimated probabilities for the witness families
already measured.
When the number of qubit pairs measured per witness family is large ($10^4$
pairs suffice in practice), $\rho_{\mathrm{true}}$ is very likely contained in
the ML set.  
Then, if there is no separable state in the ML set, we can conclude that
$\rho_{\mathrm{true}}$ is entangled. 
For finite data, this conclusion is correct within a certain error margin,
as is the case for all conclusions drawn from entanglement witness 
measurements. 
For this non-separability check, we compute the maximal values of the
likelihood for both the entire state space ($\mathcal{L}_{\mathrm{max}}$) and the
entire space of separable states ($\mathcal{L}^{\mathrm{sep}}_{\mathrm{max}}$) 
\cite{Rehacek+Hradil:2003}. 
If we find that
${\mathcal{L}_{\mathrm{max}}>\mathcal{L}^{\mathrm{sep}}_{\mathrm{max}}}$, 
we infer that $\rho_{\mathrm{true}}$ is entangled. 
To further economize the adaptive scheme, this check is performed before
looking for the unmeasured witness family with the smallest value of
$\mathcal{S}$.

\begin{figure}[t!]
\centering
\includegraphics{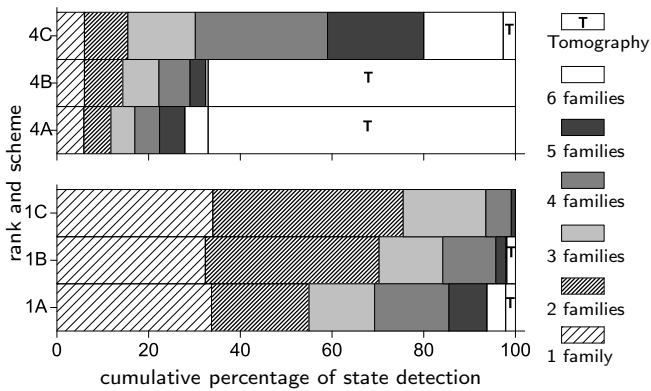}
\caption{\label{fig:histo}%
Simulation results on the measurement of the set of six informationally
complete entanglement witness families for $10^4$ randomly chosen two-qubit
entangled states: pure states (bottom) and full-rank mixed states (top). 
The cumulative histograms compare between measurements performed with scheme
A, scheme B, and scheme C.} 
\end{figure}

\heading{Simulations}
To investigate the efficiencies of the three schemes, we perform computer
simulations of witness-family measurements with both pure and full-rank mixed
two-qubit entangled states; see the Appendix
for technical details on the generation of entangled states. Figure \ref{fig:histo} shows the histograms that summarize the cumulative
distribution in the percentage of entangled states detected against the number
of witness families needed using schemes A, B and C. 
We observe that the average number of families is largest for scheme A and
smallest for scheme~C.
For instance, with only three witness families measured, 
scheme C detects about 95\% of the rank-1
entangled states whereas scheme A will need the measurement of five
families to reach the same detection rate. 
When using either scheme A or scheme B, about $2\%$ of the random pure states
and about $67\%$ of the full-rank mixed states are undetected by the six
families without performing full tomography.  
The additional separability check in scheme C reduces the percentage
of undetected pure states to virtually zero, and one needs no more than five
witness families to detect entanglement for the rest of the pure states. 
The improvement is even more dramatic for the mixed states, with a reduction
from about $67\%$ to about $2.7\%$. 
We also observe that the mean number of witness families needed to detect
entanglement for mixed states is larger than that for pure states. 
This is as expected, since mixed states generally have weaker entanglement
and are, therefore, harder to detect.  

\heading{Experiment} 
We experimentally test the entanglement detection and tomographic scheme with
three classes of states of different ranks. 
The first class of states are the pure states
${\rho^{(1)}_{\mathrm{true}}=\ket{\vartheta}\bra{\vartheta}}$, with
${\ket{\vartheta}=\ket{00}\sin\vartheta+\ket{11}\cos\vartheta}$ for
${0<\vartheta<\pi}$, ${\vartheta\neq\pi/2}$. 
The second class of states are rank-two states of the form 
${\rho^{(2)}_{\mathrm{true}}=\ket{\Phi_+}\mu\bra{\Phi_+}%
+\ket{\Phi_-}(1-\mu)\bra{\Phi_-}}$
for ${0\leq\mu\leq1}$ and ${\mu\neq1/2}$,
where ${\ket{\Phi_{\pm}}=(\ket{00}\pm\ket{11})/\sqrt{2}}$. 
The third class of states are the Werner states
${\rho^{(3)}_{\mathrm{true}}=\ket{\Psi_-}\lambda\bra{\Psi_-}+(1-\lambda)/4}$ for
${1/3<\lambda\leq1}$.
The experiment uses the polarization qubits of a down-converted photon pair at
$810\,\mathrm{nm}$ with, for example, ket $\ket{10}$ standing for
the photon in mode~1 horizontally
polarized and the photon in mode~2 vertically polarized.

\begin{figure}
\centering
\includegraphics[width=200pt]{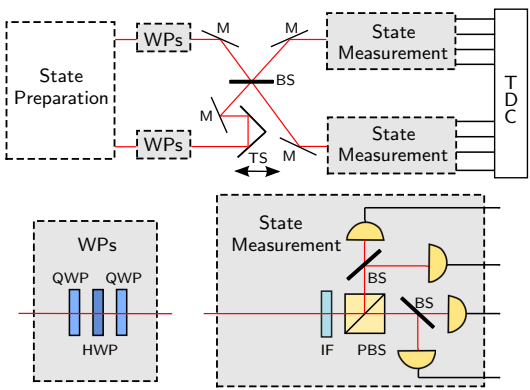}
\caption{\label{fig:exp}%
(color online)
Experimental set-up. 
The polarization-entangled two-photon states are prepared by the method
described in Ref.~\cite{Dai+4:2013}.
Upon emerging from the source, the two photons are guided with mirrors (M) to
interfere at a 50:50 beam splitter (BS), with the temporal overlap controlled 
by a translation stage (TS).
After passing through interference filters (IF), the photons are sorted by
polarizing beam splitters (PBS) and registered by one of the
photo-detectors, four on each side.
The detector outputs are addressed to a time-to-digital converter (TDC), 
and coincidences between counts of any two detectors are recorded.
Two sets of wave plates (WPs), each composed of a half-wave plate (HWP)
and two quarter-wave plates (QWP), implement the polarization
changes that correspond to the unitary operators of Table~\ref{tbl:six}.}
\end{figure}

The experimental set-up is shown in Fig.~\ref{fig:exp}. The rank-one states and rank-two states are produced 
as described in Refs.~\cite{Kwiat+4:1999,Dai+4:2013}.
The rank-four states are produced by adding a controlled admixture of 
white noise to the singlet state by varying the coincidence time window of the
detection electronics~\cite{Ling+3:2006}. 

\begin{figure}
\centering\includegraphics{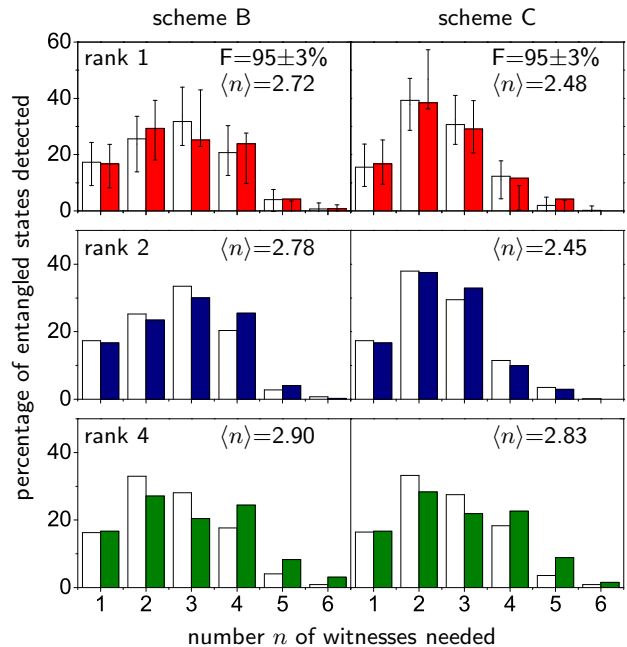}
\caption{\label{fig:data}%
(color online)
A comparison of schemes B (left column) and C (right column) for rank-one states
(top row), rank-two states (middle row), and rank-four states (bottom row). 
The histograms report the percentage of entangled states detected against the
number $n$ of witness families needed \emph{without} performing state
estimation; $\langle n\rangle$ is the average value.  
Both the simulation data (left empty bars) and the experimental data (right
full bars) show that, for the three kinds of quantum states considered, 
scheme C provides further improvement over scheme B: 
It requires fewer families on average and the distributions are narrower. 
The similarity of the two histograms for the rank-one states is confirmed by
their large fidelity $F$; similar values are obtained for the other
histograms.
--- Here, the simulation uses only states of the kind generated by the
state preparation in the set-up of Fig.~\ref{fig:exp}, whereas no such
restriction applies to the randomly-chosen states for Fig.~\ref{fig:histo}. 
} 
\end{figure}

Owing to unavoidable imperfections, the actual states emitted by the
source are not the ideal  $\rho^{(1,2,3)}_{\mathrm{true}}$ stated above but
full-rank approximations of them.  
For example, as reported in Ref.~\cite{Dai+4:2013}, fidelities above $97\%$ are
consistently achieved for the rank-two states, and the experimental rank-one 
and rank-four states are of similar quality.
The simulations for Fig.~\ref{fig:data} use the ideal states.

The signal and idler photons are directed to a Hong-Ou-Mandel 
(HOM)~\cite{Hong+2:87}
interferometer with a 50:50 beam splitter (BS).
On the way from the source to the BS, the photons pass through sets of wave
plates (WPs) that change the polarization in accordance with one of the six
local unitary transformations of Table~\ref{tbl:six}; see the Appendix
for details of the HOM calibration and settings of the WPs.
In each output port of the interferometer,
the photons pass through an interference filter with a central 
wavelength at $810\,\mathrm{nm}$ and a
full width at half maximum of $10\,\mathrm{nm}$
and are then sorted by a polarizing beam splitter (PBS). 
To discriminate between one-photon and two-photon events, another 50:50 BS
is installed into each output port of the PBS, and eight
single-photon avalanche photodiodes detect the photons.   
A time-to-digital converter (TDC) records
the arrival times of the photons, and coincidences between any two of the
eight detectors are obtained from the analysis of the time stamp record of the 
TDC. 
For each rank, we studied 21 different states.

\heading{Results} 
We set the source to a particular state and performed witness-family
measurements for one minute per family, 
and so measured about $10^4$ photon pairs for each state and family. 
The data were analyzed for all three schemes.

We also performed simulations for the three different classes of two-qubit
states. 
For each class, data from $10^4$ photon pairs (per witness family) for
$10^3$ states were simulated with Monte Carlo techniques, for the
schemes A, B, and~C. 
As expected, the simulation showed that for measurements done with the random
order of scheme A, the number of families needed for entanglement detection 
is distributed almost evenly from one to six, with a mean of about $3.5$, 
and fewer families need to be measured in the adaptive schemes.

This is confirmed by the experiment. 
For the three classes of states, only one of the witness families gives a
conclusive result.  
Hence, for scheme A, the number of
families needed for entanglement detection is equally likely to be one to
six.  The data for schemes B and C were analyzed as explained in 
the Appendix and the results are
shown in Fig.~\ref{fig:data}, where
one observes a significant improvement over the non-adaptive scheme~A, which needs about 3.5 families on average.
One rarely needs more than four witness families to detect the
entanglement in these states; full tomography is never necessary since all
rank-one, rank-two, and rank-four states are detected by one of the witness
families. 
 
The striking similarity between the results of the simulation and the
experiment indicates that there are no significant systematic errors in the
experimental data.
For the rank-one states, the fidelity $F=\sum_j(p_jp_j')^{1/2}$ 
between two probability distributions $\{p_j\}$ and $\{p_j'\}$ 
compares the histograms from simulation and experiment, and the large value of
$F$ is reassuring.
The error bars for the rank-one states are obtained by bootstrapping the
actual data one hundred times;
these error bars show the variation in the histograms that repeated
measurements of this kind would display. 
The fidelity values and the error bars for the other histograms are of similar
sizes and not displayed. We also
collected IC data even if the state is known to be entangled before
all six witness-family measurements are done so as to check the reliability
of the tomography. The witness-family measurements indeed enable reliable quantum
state tomography; see the Appendix
for details.

\heading{Conclusions}
We performed an experiment to verify the witness-family-based
entanglement detection scheme introduced in Ref.~\cite{Zhu+2:2010}. 
In going beyond that proposal, we also introduced adaptive schemes that use
the information acquired in previous measurements to reduce the average number
of witness families that need to be measured.
A few-witness way of detecting entanglement for photon-polarization qubits
was thus demonstrated. 
Further, we showed that the witness-family measurements enable reliable quantum
state tomography. 
With the necessary changes and within the limits set by what is experimentally
feasible, 
the witness-family approach is also applicable to qubits of other physical
kinds than photon polarization and to higher-dimensional or multi-partite
systems.

\heading{Acknowledgments}
We are grateful for the insightful discussions with Jaroslav \v{R}eh\'a\v{c}ek
and Huangjun Zhu.  
We thank Norbert L\"utkenhaus for bringing Ref.~\cite{Guhne+1:2006} to our
attention.
This work is supported by the National Research Foundation and the Ministry of
Education, Singapore, and is co-financed by the European Social Fund and the
state budget of the Czech Republic, project No.\ CZ.1.07/2.3.00/30.0004
(POST-UP).

\bigskip

\centerline{\textbf{APPENDIX}}
In this appendix, we comment on related work, discuss more general adaptive
schemes, provide some technical details of the experiment 
and present our tomographic results.

\section{The experiment by Barbieri \textit{et al.}}  
Upon denoting the Pauli operators of the $j$th qubit by $X_j$, $Y_j$, and
$Z_j$, the witness of Eq.~(\ref{eq:Walfa}) is
\begin{eqnarray}\label{eq:SM1}
  W^{(\alpha)}&=&\frac{1}{4}\bigl[1+(Z_1+Z_2)\cos\alpha+Z_1Z_2
\nonumber\\&&\phantom{\frac{1}{4}\bigl[}+(X_1X_2+Y_1Y_2)\sin\alpha\bigr]\,.
\end{eqnarray}
In the experiment of Barbieri \textit{et al.}\ \cite{Barbieri+5:2003}
the two polarized photons are measured separately and, therefore, a direct
measurement of $W^{(\alpha)}$ is not possible.
Instead, the expectation value of $W^{(\pi/2)}$ is inferred from three auxiliary
measurements that detect the common eigenstates (i) of $X_1$ and $X_2$, 
(ii) of $Y_1$ and $Y_2$, and (iii) of $Z_1$ and $Z_2$.

Together, the data acquired in these measurements establish the nine
expectation values
\begin{equation}\label{eq:SM2}
\begin{array}{c@{\,,\ }c@{\,,\ }c}
  \expect{X_1}&\expect{X_2}&\expect{X_1X_2}\,,\nonumber\\
  \expect{Y_1}&\expect{Y_2}&\expect{Y_1Y_2}\,,\nonumber\\
  \expect{Z_1}&\expect{Z_2}&\expect{Z_1Z_2}\,,
\end{array}  
\end{equation}
of which only the three values in the right column are used for
$\bexpect{W^{(\pi/2)}}$.
The other six values are not exploited.
While this is wasteful, it does not matter in the context of
Ref.~\cite{Barbieri+5:2003} where one knows beforehand that the unknown state
is of the form
\begin{equation}\label{eq:SM3}
  \rho=\frac{1}{4}\bigl[1-p(X_1X_2+Y_1Y_2+Z_1Z_2)\bigr]
\end{equation}
with ${0\leq p\leq1}$.
The value of parameter $p$ is then provided by ${\bexpect{W^{(\pi/2)}}=(1-3p)/4}$. 

All witnesses in the first and the second family of Table~\ref{tbl:six}
are made available by the five expectation values in the bottom row and the
right column in (\ref{eq:SM2}).
It follows that the inequality
\begin{eqnarray}\label{eq:SM4}
  1+\expect{Z_1Z_2}^2&\geq&2\bigl|\expect{X_1X_2}\expect{Y_1Y_2}-\expect{Z_1Z_2}
+\expect{Z_1}\expect{Z_2}\bigr|\nonumber\\
&&\mbox{}+\expect{X_1X_2}^2+\expect{Y_1Y_2}^2
+\expect{Z_1}^2+\expect{Z_2}^2\nonumber\\
\end{eqnarray}
holds for all separable states.
There are two more inequalities of the same structure for the expectation
values of the first and the second row, respectively, and the third column in
(\ref{eq:SM2}).

If all three inequalities are obeyed, further measurements are needed.
They would detect the common eigenstates of $X_1$ and $Y_2$, of $Y_1$ and
$X_2$, and so forth. 
Each such measurement gives the expectation values of two single-qubit
operators already contained in (\ref{eq:SM2}) and adds one new two-qubit
expectation value to the list.
More inequalities analogous to (\ref{eq:SM4}) become available in the course.
Eventually, when all six two-qubit expectation values that are missing in
(\ref{eq:SM2}) are determined, full tomography is achieved.

Clearly, the sequence in which the fourth, fifth, \dots\ ninth measurements
are carried out, can be optimized by an adaptive strategy.
There is also the option of checking, at each stage, whether there are
separable states in the convex set of maximum-likelihood estimators, and
inferring that the unknown state is entangled if there are none.   

In summary, the measurement scheme of Barbieri \textit{et al.}\ can be used
for an indirect measurement of witness families.  
When state reconstruction is necessary, this indirect measurement needs nine
settings, each providing three expectation values of the  
fifteen independent ones; each of the six single-qubit expectation values is
determined thrice. 
By contrast, the direct measurement achieves full tomography with six
settings, whereby three of the nine two-qubit expectation values are
determined twice (see Table II in Ref.~\cite{Zhu+2:2010}). 
Although one could conclude that the direct measurement is less wasteful and
should be preferred over the indirect measurement, one must remember that the
direct measurement 
is not feasible when the two polarized photons are at different locations. 
Then, the Barbieri \textit{et al.}\ scheme does the job.

\section{The nonlinear witnesses of G\"uhne and L\"utkenhaus}

The inequality ${\bexpect{W^{(\alpha)}}\geq0}$ holds for all separable states;
this is, of course, the witness property.
In Ref.~\cite{Guhne+1:2006}, G\"uhne and L\"utkenhaus show that the null
bound can be replaced by various state-dependent positive bounds,
\begin{equation}\label{eq:SM5}
\bexpect{W^{(\alpha)}}\geq\bigl|\bexpect{G^{(\alpha)}}\bigr|^2\,,
\end{equation}
where, for example, the non-Hermitian operators
\begin{eqnarray}\label{eq:SM6}
  G_1^{(\alpha)}
   &=&\frac{1}{\sqrt{8}}\sin\Bigl(\frac{\alpha}{2}+\frac{\pi}{4}\Bigr)
               (1+X_1X_2+Y_1Y_2+Z_1Z_2)
\nonumber\\ &&\mbox{}
+\frac{1}{\sqrt{8}}\cos\Bigl(\frac{\alpha}{2}+\frac{\pi}{4}\Bigr)
(Z_1+Z_2+\I X_1Y_2-\I Y_1X_2)\nonumber\\
\end{eqnarray}
or
\begin{eqnarray}\label{eq:SM7}
  G_2^{(\alpha)}
    &=&\frac{1}{\sqrt{8}}\sin\Bigl(\frac{\alpha}{2}+\frac{\pi}{4}\Bigr)
               (X_1+X_2+\I Y_1Z_2-\I Z_1Y_2)
\nonumber\\ &&\mbox{}
+\frac{1}{\sqrt{8}}\cos\Bigl(\frac{\alpha}{2}+\frac{\pi}{4}\Bigr)
(X_1Z_2+Z_1X_2+\I Y_1-\I Y_2)
\nonumber\\
\end{eqnarray}
are possible choices for $G^{(\alpha)}$.
The lower bounds in Eq.~(\ref{eq:SM5}) are second-degree polynomials of
one-qubit and two-qubit expectation values; there are also lower bounds that
are polynomials of fourth or higher order.

We can evaluate $\bexpect{G_1^{(\alpha)}}$ as soon as the three witness
families 1, 3 (or 4), and 5 (or 6) have been measured; similarly,
$\bexpect{G_2^{(\alpha)}}$ is available after measuring the four families 3--6.
This illustrates that partial tomography is needed before the more stringent
lower bounds of Eq.~(\ref{eq:SM5}) are at hand.

Nevertheless, it could be interesting to exploit criteria of this kind (with
parameter $\alpha$ optimized) for a possible further reduction of the number
of witness families that need to be measured before one can conclude that the
unknown state is entangled.
This is unexplored territory.

\section{General adaptive schemes}
The adaptive schemes B and C that speed up the entanglement detection select
the next witness family from the six pre-chosen families specified by
the unitary operators in Table~\ref{tbl:six}. 
If, instead, one selects also from other families than the six pre-chosen
ones, such a more general adaptive scheme might be more efficient, in the
sense that fewer families need to be measured on average before one can
conclude that the unknown state is entangled.
Thereby, the selection is still done by opting for the family which
is expected to give the smallest value of $\mathcal{S}$ upon measurement.
It turns out that the more general adaptive schemes are not worth the trouble. 
We justify this remark by a study of the generalizations of schemes B and C.

\heading{Scheme B'} 
As in scheme B, we calculate the MLME estimator $\rho_{\textsc{mlme}}$ and
exploit its properties when choosing the next family to be measured. 
If $\rho_{\textsc{mlme}}$ is entangled,
$\rho_{\textsc{mlme}}^{\mathrm{T}_2}$ has one negative eigenvalue, and 
the eigenket $\ket{\phi}$ to this eigenvalue is entangled. 
The next witness family is then the one obtained from
${W=\bigl(\ket{\phi}\bra{\phi}\bigl)^{\mathrm{T}_2}}$ because 
this family is best for detecting the entanglement of
$\rho_{\textsc{mlme}}$. 
Since $\rho_{\textsc{mlme}}$ is the current best guess for the unknown state
$\rho_{\mathrm{true}}$, this family has also a good chance of detecting
entanglement in $\rho_{\mathrm{true}}$.
If the MLME estimator is separable, however, we proceed as in scheme B.

\heading{Scheme C'} 
On top of scheme B', the separability check of scheme C is implemented.

\begin{figure}
\centering
\includegraphics[width=240pt]{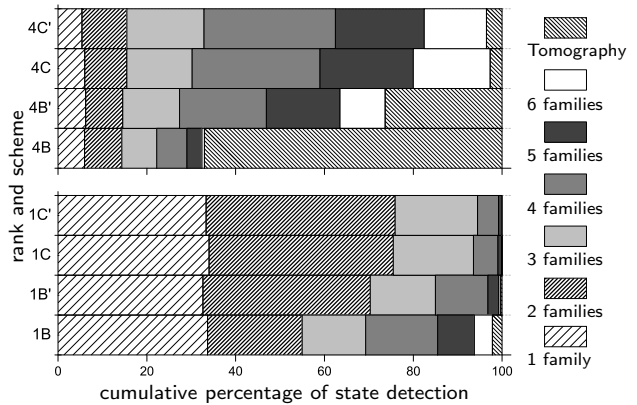}
\caption{\label{sfig:histo}%
Simulation results for $10^4$ randomly chosen two-qubit
entangled states: pure states (bottom) and full-rank mixed states (top).   
The cumulative histograms compare between adaptive measurements performed with
the six pre-chosen families of Table~\ref{tbl:six} (schemes B and C)  
and with six arbitrary families (schemes B' and C').} 
\end{figure}

\heading{Simulations}
We investigate the general adaptive schemes by performing computer simulations
for pure and full-rank states and constructing histograms analogous to those
in Fig.~\ref{fig:histo}. 
As can be seen from the results in Fig.~\ref{sfig:histo}, scheme B' improves
over scheme B.  
For both rank-one and rank-four states, the cumulative percentage of states
detected is higher. 
The percentage of undetected rank-four states after six families drops from
about $67\%$ to about $25\%$. 
While this improvement is substantial, it pales in comparison with the
dramatic reduction to about $2.7\%$ when using scheme C. 
Therefore, when aiming at the most efficient way of detecting entanglement, we
have to employ either scheme C or scheme C'. 

Now, as we learn from Fig.~\ref{sfig:histo}, scheme C' is only slightly
better than scheme C --- if at all.
For, the somewhat smaller proportion of rank-four states detected  by the first,
randomly chosen, family and the somewhat larger proportion of states requiring
tomography are surely resulting from statistical fluctuations in the
simulation. 
The other differences between the histograms for schemes C and C' are of
similar size. 
Accordingly, there is no evidence that scheme C' is worth the trouble of its
implementation, which requires that the settings of the WPs are calculated and
updated in every step for every state.
On the other hand, the data obtained in scheme C are optimal for the
tomographic reconstruction of $\rho_{\mathrm{true}}$. 
In summary, then, scheme C serves all purposes very well.   

\section{Generation of random entangled states}
For all the simulations, we generate a unitarily invariant ensemble of random entangled states by the
procedure of Ref.~\cite{Zyczkowski+1:01}: 
For each random state, we first compose an auxiliary matrix 
$\mathcal{A}$ of dimensions $1\times4$ (for a pure state) or $4\times4$
(for a full-rank state), with the random complex entries chosen from a normal
distribution with zero mean and unit variance;
then, the $4\times4$ matrix representing the random state is 
$\mathcal{A}^\dagger\mathcal{A}/\tr{\mathcal{A}^\dagger\mathcal{A}}$. 
Most of the random states generated this way are entangled, since there are
many more entangled states in such an ensemble than separable states;
we check the concurrence of each state 
to ensure that only entangled states are used in the simulation.

\section{Details of the experiment}

\heading{Wave plate settings}
Each of the local unitary operators of Table~\ref{tbl:six} is implemented by a
HWP between two QWPs, 
\begin{equation}\label{eq:sandwich}
  U=U_\mathrm{QWP}(\alpha)U_\mathrm{HWP}(\beta)U_\mathrm{QWP}(\gamma)\,,
\end{equation}
where the QWP with angle $\gamma$ is the first in the sequence; 
see Fig.~\ref{fig:exp}.
The matrix representation for a HWP, apart from an irrelevant global phase
factor, is (see, e.g., \cite{Englert+2:01}) 
\begin{eqnarray}
U_\mathrm{HWP}(\theta)&=&
\left(\begin{array}{c@{\enskip}c}
\ket{\textsc{v}}&\ket{\textsc{h}}
\end{array}\right)
\left(\begin{array}{c@{\enskip}c}
\cos(2\theta) & \sin(2\theta) \\ \sin(2\theta)  & -\cos(2\theta)
\end{array}\right)
\left(\begin{array}{c}
\bra{\textsc{v}}\\\bra{\textsc{h}}
\end{array}\right)\nonumber\\
&\repr&\left(\begin{array}{c@{\enskip}c}
\cos(2\theta) & \sin(2\theta) \\ \sin(2\theta)  & -\cos(2\theta)
\end{array}\right),
\end{eqnarray}
where $\theta$ is the angle between its major axis and the vertical
direction,
and we have 
\begin{equation}
U_\mathrm{QWP}(\theta)\repr\frac{1}{\sqrt{2}}
\left(\begin{array}{c@{\quad}c}
1-\I \cos(2\theta) & -\I \sin(2\theta)\\ 
-\I \sin(2\theta) & 1+\I \cos(2\theta) \end{array}\right)
\end{equation}
for a QWP.
Further, the matrices for the Pauli operator $X$ that permutes
$\ket{\textsc{v}}$ and $\ket{\textsc{h}}$ and the Clifford operator $C$ that
permutes the three Pauli operators cyclically are  
\begin{equation}
X\repr\left(\begin{array}{c@{\enskip}c}
0 & 1 \\1 & 0 \end{array}\right) \quad\textrm{and}\quad
C\repr\frac{1}{\sqrt{2}}\left(\begin{array}{c@{\enskip}c}
1 & -\I  \\1 & \ \I  \end{array}\right),
\end{equation}
respectively.

\begin{table}
\caption{\label{stbl:six}%
Wave plate settings for the unitary operators of Table~\ref{tbl:six}.
The angles $\alpha$, $\beta$, and $\gamma$ are such that the corresponding $U$
is obtained from Eq.~(\ref{eq:sandwich}).}
\centering
    \begin{tabular}{c@{\qquad}c@{\qquad}c@{\qquad}c}
    \hline \hline
   $U$ & $\alpha$ & $\beta$ & $\gamma$\\  \hline
    $\mathbf{1}$ & 0  & 0 & 0 \\ 
    $X$ & 0 & $\pi/4$ & 0 \\
    $C$ & 0 & $\pi/4$ & $-\pi/4$\\  
    $C^\dagger$ & $-\pi/4$  & 0 & $0$ \\ 
    $XC$ & $0$ & 0 & $-\pi/4$\\
    $XC^\dagger$ & $\pi/4$ & 0 & 0\\
    \hline \hline
    \end{tabular}
\qquad
\end{table}

The angles $\alpha$, $\beta$, and $\gamma$, for which the various $U$s are
realized, are reported in Table~\ref{stbl:six}.
As an example, we consider
\begin{equation}
U=XC\, \, \widehat{=}\,\frac{1}{\sqrt{2}}
\left(\begin{array}{c@{\enskip}c}
0 & 1\\ 1 & 0
\end{array}\right)
\left(\begin{array}{c@{\enskip}c}
1 & -\I  \\1 & \I  
\end{array}\right)
=\frac{1}{\sqrt{2}}
\left(\begin{array}{c@{\enskip}c}
1 & \I  \\ 1 & -\I  
\end{array}\right)
\end{equation}
and verify that $\alpha = 0$, $\beta = 0$, and $\gamma = -\pi/4$ are
correct choices.
Indeed, they are:
\begin{eqnarray}
&&U_\mathrm{QWP}(0)U_\mathrm{HWP}(0)U_\mathrm{QWP}(-\pi/4)
\nonumber\\
&\widehat{=}&\frac{1-\I }{\sqrt{2}}
\left(\begin{array}{c@{\enskip}c}
1& 0 \\ 0 &\I   
\end{array}\right)
\left(\begin{array}{c@{\enskip}c}
1 & 0 \\ 0 & -1 
\end{array}\right)\frac{1}{\sqrt{2}}
\left(\begin{array}{c@{\enskip}c} 
1 & \I  \\ \I  & 1 
\end{array}\right)\nonumber\\
&=&
\Exp{-\I\pi/4}
\frac{1}{\sqrt{2}}
\left(\begin{array}{c@{\enskip}c}
1 & \I  \\ 1& -\I  
\end{array}\right),
\end{eqnarray}
since the global phase factor is irrelevant.

\begin{figure}
\centering
\includegraphics[width=230pt]{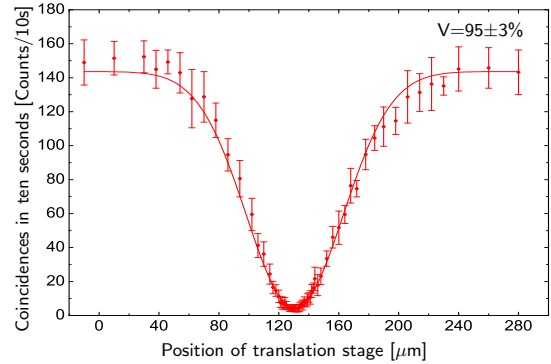}
\caption{\label{sfig:HOMdip}
An example of a HOM dip obtained in our experiment for 
the state $\ket{\textsc{hh}}\bra{\textsc{hh}}$. 
The visibility, $V$, of the HOM dip above is $95\pm3\%$; other HOM dips
observed for different polarization states are similar to this one.
}
\end{figure}

\heading{HOM interferometer}
To implement the most direct measurement of the witness bases, we make use of
a HOM interferometer. 
For its optimization, we first remove the BSs in the output ports of 
the PBSs; see Fig.~\ref{fig:exp}. 
Next, we address the outputs of the two detectors in the transmission ports of
the two PBSs to a coincidence unit, where we record the coincidence rate for
the input two-photon state $\ket{\textsc{hh}}$. 
A translation stage with a step size of $500\,\mathrm{nm}$ is used to control
the temporal overlap between the photons, and 
the spatial overlap is controlled by adjusting the mirrors that direct the two
beams to interfere on the BS. 
The interferometer is optimized at where the coincidence rate is minimal, 
as shown in Fig.~\ref{sfig:HOMdip}. 
The visibility of the HOM dip is $95\pm3\%$. 
Imperfections in the BS ratios, the WPs, and the polarization controllers
limit the maximum experimental achievable visibility of the dip.
Nevertheless, the visibilities of the HOM dips obtained in our experiment
exceed $90\%$ for all the data collected for the construction of the
histograms shown in Fig.~\ref{fig:data}.

\heading{Data Analysis}
Here we illustrate how data were analyzed to obtain Fig.~3 in the main text.
For the witness family measurement, the first family is chosen at random among the six families. 
If the measurement of this family gives a negative value of $\mathcal{S}$,
the state is detected to be entangled and no further measurement is needed.
If the result is inconclusive, then one uses the adaptive scheme to choose
the next family, until a conclusive result is obtained. 
The result is $n$, the number of witness families that have to be measured in
order to detect the entanglement; see Table~\ref{tbl:Expl} for illustrations.
\begin{table}
\caption{\label{tbl:Expl}%
Examples demonstrating how Fig.~\ref{fig:data} is derived. 
The first family is chosen at random among the six families. 
If the measurement of this family gives a negative value of $\mathcal{S}$,
the state is detected to be entangled and no further measurement is needed.
If the result is inconclusive, then one uses the adaptive scheme to choose
the next family, until a conclusive result is obtained. 
The result is $n$, the number of witness families that have to be measured in
order to detect the entanglement.}
\centering
\begin{tabular}{c@{\quad}ccc@{\quad}c}\hline\hline
State &\multicolumn{3}{c}{%
{Families} and their $\mathcal{S}$ values\mbox{\quad}}& 
$n$ \\\hline\rule{0pt}{12pt}%
$\rho^{(2)}_{\mathrm{true}}$ & family~3 & family~4 & family~2 & 3\\
 $\mu=0.15$ & $0.46\pm0.03$  
 & $0.06\pm0.02$ & $-0.23\pm0.04$ &  
\\[2ex]
$\rho^{(3)}_{\mathrm{true}}$ & family~1 & --- & --- & 1\\
$\lambda=1$    & $-0.83\pm0.03$ & & &\\[2ex]
$\rho^{(1)}_{\mathrm{true}}$ & family~4 & family~2 & --- & 2\\
 $\vartheta=\pi/4$    & $0.21\pm0.03$ & $-0.87\pm0.03$ & & \\
\hline\hline  
\end{tabular}
\end{table}

\section{Tomographic results}
For the three classes of states, all entangled states are successfully
detected without the need to perform full tomography. 
There are, however, other entangled states that would escape detection, and
all separable states can only give inconclusive results.
Regarding entangled states, we recall that
about $2\%$ of the random pure states and about $67\%$ of the random mixed
states are not detected by the six witness families 
without the separability check of scheme C. 
Hence, to confirm the efficiency and accuracy of the tomographic scheme, we
collect IC data even if the state is known to be entangled before
all six witness-family measurements are done. 
Using the technique of ML estimation \cite{Banaszek+3:00,Rehacek+3:07}, 
we then infer the state
from the data and calculate the fidelity $\displaystyle%
\mathrm{tr}\bigl\{|\sqrt{\rho_{\mathrm{true}}}\sqrt{\rho_{\mathrm{est}}}|\bigr\}$
between the true and the estimated state. 
The average fidelities are $98.3\pm0.7\%$, $97.4\pm1.4\%$, and
$98.7\pm1.1\%$ for the respective true states of rank one, two, and four. 
Indeed, if all the six family measurements fail to detect the entanglement in
the state, one can use the tomographic information to reliably reconstruct the
unknown input state and then determine its separability numerically.

\end{document}